\documentclass[twocolumn,aps,preprintnumbers,bibnotes10pt,superscriptaddress,longbibliography]{revtex4-1}

\usepackage{amsmath}
\usepackage{graphicx}
\usepackage[normalem]{ulem}
\usepackage{dcolumn}
\usepackage{epsfig}
\usepackage{bm}
\usepackage{array}
\usepackage[english]{babel}
\usepackage{hyperref}
\usepackage{algorithm,algorithmic}

\hypersetup{
colorlinks=true,
citecolor=blue,
linkcolor=red,
urlcolor=blue
 , bookmarks=true,
 pdfmenubar=true
}

\usepackage{hyperref}
\usepackage{graphicx}
\usepackage{amsmath}
\usepackage{amsfonts}
\usepackage{amssymb}
\usepackage{xcolor}
\usepackage{bbm}
\usepackage{calrsfs}
\usepackage{dutchcal}
\usepackage{amsthm}

\AtBeginDocument{%
    \newwrite\bibnotes
    \def\bibnotesext{Notes.bib}
    \immediate\openout\bibnotes=\jobname\bibnotesext
    \immediate\write\bibnotes{@CONTROL{REVTEX41Control}}
    \immediate\write\bibnotes{@CONTROL{%
    apsrev41Control,author="08",editor="1",pages="1",title="0",year="1"}}
     \if@filesw
     \immediate\write\@auxout{\string\citation{apsrev41Control}}%
    \fi
}%

\newcommand{\bra}[1]{\ensuremath{\langle#1|}}
\newcommand{\ket}[1]{\ensuremath{|#1\rangle}}
\newcommand{\Eins}{\ensuremath{\mathbbm 1}}

\newcommand{\bs}[1]{\boldsymbol{#1}}

\newcommand{\be}{\begin{equation}}
\newcommand{\ee}{\end{equation}}
\newcommand{\ud}{\mathrm{d}}
\newcommand{\beq}{\begin{eqnarray}}
\newcommand{\eeq}{\end{eqnarray}}

\begin{document}

\title{Bayesian Quantum Multiphase Estimation Algorithm}
\author{Valentin~Gebhart}
\affiliation{QSTAR, INO-CNR and LENS, Largo Enrico Fermi 2, 50125 Firenze, Italy}
\affiliation{Universit\`a degli Studi di Napoli ”Federico II”, Via Cinthia 21, 80126 Napoli, Italy}

\author{Augusto~Smerzi}
\affiliation{QSTAR, INO-CNR and LENS, Largo Enrico Fermi 2, 50125 Firenze, Italy}

\author{Luca~Pezz\`e}
\affiliation{QSTAR, INO-CNR and LENS, Largo Enrico Fermi 2, 50125 Firenze, Italy}

\begin{abstract}
Quantum phase estimation (QPE) is the key subroutine of several quantum computing algorithms
as well as a central ingredient in quantum computational chemistry and quantum simulation.
While QPE strategies have focused on the estimation of a single phase, applications to the simultaneous estimation of several phases may bring substantial advantages; for instance, in the presence of spatial or temporal constraints.
In this work, we study a Bayesian algorithm for the parallel (simultaneous) estimation of multiple arbitrary phases.
The protocol gives access to correlations in the Bayesian multi-phase distribution resulting in  covariance matrix elements scaling as $O(N_T^{-2})$, with respect to the total number of quantum resources $N_T$.
The parallel estimation allows to surpass the sensitivity of sequential single-phase estimation strategies for optimal linear combinations of phases. 
Furthermore, the algorithm proves a certain noise resilience and can be implemented using single photons and standard optical elements in currently accessible experiments.
\end{abstract}

\maketitle

\section{Introduction}

Quantum phase estimation (QPE) provides a link between quantum computation, simulation, and  metrology.
In fact, several quantum technology tasks can be cast as the estimation of an unknown eigenphase $\theta \in [0,2\pi]$ of a unitary matrix $U$, $U \ket{u} = e^{i \theta}\ket{u}$, where $\ket{u}$ is 
the corresponding eigenstate.
The problem can be efficiently solved using QPE algorithms~\cite{Kitaev1996,Kitaev2002,Cleve1998}. QPE is the key subroutine of quantum algorithms providing exponential speedup~\cite{Nielsen2010}, 
including number factoring~\cite{Shor1999,Martin-Lopez2012,Monz2016},
 computation of molecular spectra~\cite{Abrams1999,Aspuru-Guzik2005,OMalley2016,McArdle2020}, and
quantum sampling~\cite{Temme2011}. It also finds applications in the alignment of spatial reference frames~\cite{Rudolph2003}, clock synchronization~\cite{DeBurgh2005}, frequency estimation~\cite{Kessler2014,Pezze2020}, and clock networks~\cite{Komar2014}.
In quantum metrology~\cite{Giovannetti2011,Toth2014,Pezze2018}, ideas developed for QPE algorithms have been applied to the estimation of an interferometric phase $\theta$ with Heisenberg-limited sensitivity, both locally~\cite{Pezze2007a} and 
in the full $\theta\in[0,2\pi]$ domain~\cite{Higgins2007b,Higgins2009,Berry2009,Berni2015,Daryanoosh2018,Pezze2020a}.
Experimental QPE has been implemented with single photons~\cite{Higgins2007b,Higgins2009},
including the simulation of Hamiltonian spectra on an integrated photonic device~\cite{Paesani2017,Santagati2018}, and NV centers~\cite{Bonato2016}.

So far, QPE algorithms have mainly focused on the estimation of a single eigenphase of $U$. 
Besides tailored postprocessing techniques~\cite{OBrien2019a,Somma2019},
existing multiphase estimation algorithms follow a sequential scheme where an ancilla qubit encodes a single parameter at a time and is then measured.
However, there is an urgent demand for parallel-multiphase-estimation techniques~\cite{Humphreys2013,Ragy2016,Ciampini2016,Proctor2018,Ge2018,Gessner2018b,Oh2020,Rubio2020,Gessner2020,Goldberg2020}, where 
different phases are estimated simultaneously~\cite{Polino2019,Guo2020,Valeri2020}.
Indeed, several problems of interest involve the joint estimation of multiple parameters and/or their linear combinations.
Important examples include imaging~\cite{Albarelli2020,Rehacek2017},  sensing~\cite{Baumgratz2016,Apellaniz2018,Degen2017}, and biological probing~\cite{Taylor2016,Cimini2019}.

In this paper, we extend Kitaev's iterative QPE algorithm \cite{Kitaev1996,Kitaev2002,Higgins2007b, Berry2009, Wiebe2016} to the {\it parallel} (simultaneous) estimation of $d\geq1$ eigenphases $\boldsymbol{\theta} = (\theta_1, ..., \theta_d)$ of $U$, using a finite number of resources $N_T$.
Here, $N_T$ is quantified as the total number of applications of a controlled-$U$ gate~\cite{Giovannetti2006,Higgins2007b,Berry2009}.
In general, parallel multiphase estimation requires us to access phase correlations in the output state (which are optimally captured by a Bayesian approach) while avoiding the proliferation of marginal-probability tails in the $d$-dimensional posterior phase distribution.
Our protocol achieves a sensitivity  scaling $O(N_T^{-1})$ for the simultaneous estimation of 
arbitrary vector parameters $\boldsymbol{\theta} \in [0, 2\pi]^d$, as well as for arbitrary linear combinations of the $d$ phases. 
The algorithm inherits important properties from Bayesian approaches to single-parameter estimation~\cite{Pezze2007a,Wiebe2016,Paesani2017,Bonato2016,Li2018a,Nolan2020}, such as noise resilience and the optimal use of measurement data. 
The exploitation of correlations in the $d$-dimensional Bayesian distribution allows us to
overcome the sensitivity achievable with the sequential scheme for the estimation of certain linear combinations of the phases, for fixed $N_T$. Owing to the primary relevance of QPE in quantum technologies, our methods may be applied to a variety of relevant problems and algorithms related to joint estimation of multiple (eigen)phases.

Our protocol goes beyond the usual approach to multiparameter estimation ~\cite{Humphreys2013,Ragy2016,Ciampini2016,Proctor2018,Ge2018,Gessner2018b,Oh2020,Gessner2020} based of the analysis of the quantum Fisher information matrix (QFIM) ~\cite{Helstrom1976,Holevo1982,Pezze2017}. First, the optimal measurement strategy to saturate the QFIM generally depends the specific $\boldsymbol{\theta}$, while our algorithm uses a single approach for the estimation of any $\boldsymbol{\theta} \in [0, 2\pi]^d$.
Furthermore, the (frequentist) sensitivity predicted by the QFIM is only achieved asymptotically in the number of repeated measurements, which generally hinders the scaling with respect to the total number of resources.
In particular, fixing $N_T$ allows us to clarify unambiguously under which conditions parallel strategies overcome sequential ones, thus settling a long-standing debate in the literature \cite{Humphreys2013,Ragy2016,Ciampini2016,Proctor2018,Ge2018,Gessner2018b}.

The paper is structured as follows. In Sec.~\ref{sec:alogorithm}, we present the quantum circuit and the corresponding Bayesian algorithm for the parallel estimation of  $d \geq 1$ eigenphases of a generic unitary operator $U$.
In Sec.~\ref{sec:simulation}, we discuss the performance of the algorithm in both the presence and the absence of noise. In Sec.~\ref{sec:impl}, we propose a multiround protocol for the parallel estimation of $d$ optical phases that can be implemented with single photons and standard optical elements. Finally, we conclude in Sec.~\ref{sec:conclusion}.

\section{Bayesian algorithm for parallel quantum multiphase estimation}\label{sec:alogorithm}

\subsection{Quantum circuit of the algorithm}

\begin{figure}[t]
	\center
	\includegraphics[width = 1\columnwidth]{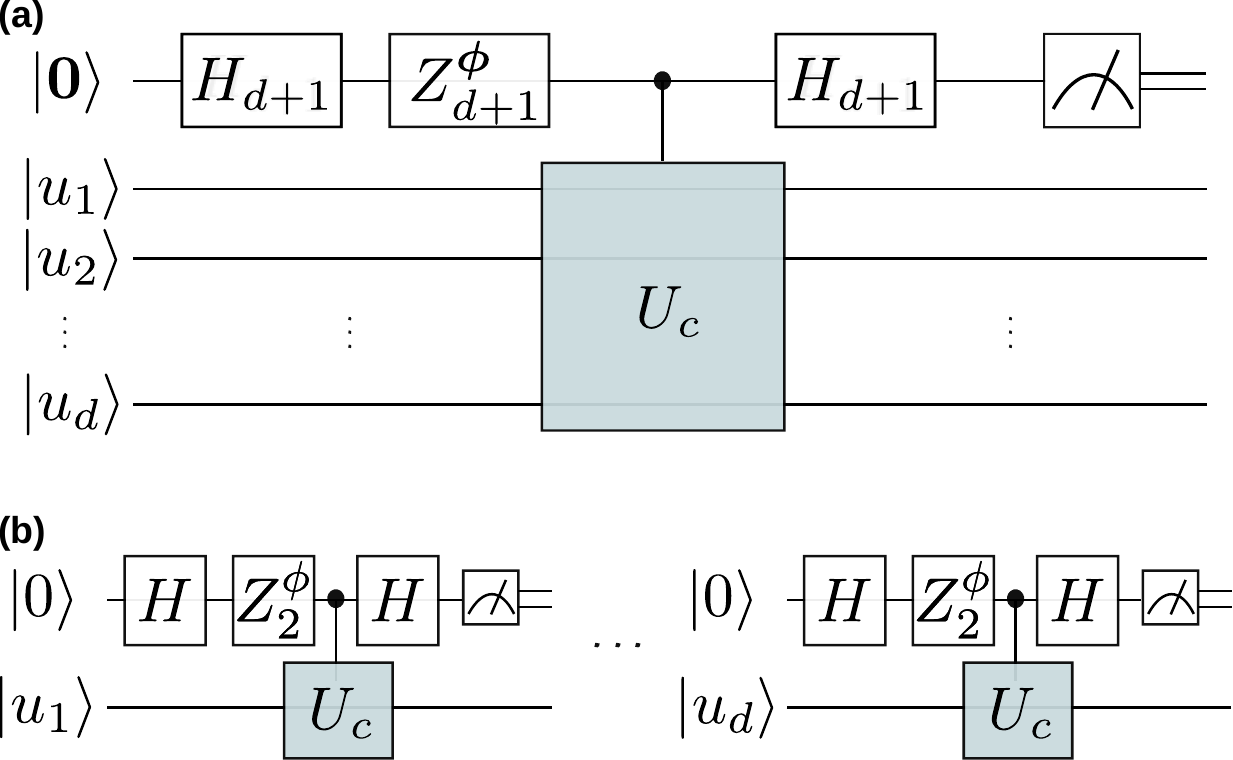}
	\caption{
	Quantum circuits for parallel (a) and sequential (b) multiphase estimation. The corresponding quantum algorithm is detailed in the text. $H_{d+1}$ and $Z_{d+1}^{\boldsymbol{\phi}}$ ($H$ and $Z_2^\phi$) are generalized (single-qubit) Hadamard and phase gates, respectively.
	$\ket{\boldsymbol{0}}$ denotes a $(d+1)$-dimensional ancilla qudit, $\ket{0}$ an ancilla qubit and $U_c$ is a controlled-$U$ gate.
	}
	\label{fig1}
\end{figure}

Our iterative multiphase estimation algorithm is based on the quantum circuit shown in Fig.~\ref{fig1}(a). 
As is common in QPE algorithms~\cite{Kitaev1996,Cleve1998,Nielsen2010,Abrams1999,Aspuru-Guzik2005}, 
we assume (i) the existence of an oracle having access to $d$ eigenstates of $U$ and 
(ii) the possibility of generating controlled-$U^{M}$ gates, with $M$ being an integer that we specify below.
The circuit consists of a register of $d$ eigenstates $\ket{u_1}, ..., \ket{u_d}$ of $U$ and a $(d+1)$-dimensional ancilla qudit.
The register is prepared in the product state $\bigotimes_{j=1}^d \ket{u_j}$.
The ancilla is initialized
in $\ket{\mathbf{0}}$
(bold for the ancilla states) and then transformed
by a $(d+1)$-dimensional generalized Hadamard gate $H_{d+1}$ defined by $\bra{\mathbf{m}} H_{d+1} \ket{\mathbf{n}}=e^{i[ 2\pi mn/(d+1)]}/\sqrt{d+1}$, where $m,n \in \{ 0,\dots,d \}$ label the computational basis states.
This gate prepares the ancilla in the superposition $(\ket{\mathbf{0}}+\dots+\ket{\mathbf{d}})/\sqrt{d+1}$.
Next, $d+1$ adjustable (known) controlled phases
$\boldsymbol{\phi} = ( \phi_0, ..., \phi_{d} )$ are imprinted on the ancilla via $Z _{d+1}(\boldsymbol{\phi})$ given by $\bra{\mathbf{m}} Z _{d+1}(\boldsymbol{\phi}) \ket{\mathbf{n}}=\delta_{mn}e
^{i\phi_m}$.
The key operation of the quantum circuit is the generalized controlled-$U$ gate 
\begin{multline} \label{CPG}
U_c=\ket{\mathbf{0}}\bra{\mathbf{0}} \otimes \Eins\otimes \cdots\otimes \Eins + \ket{\mathbf{1}}\bra{\mathbf{1}} \otimes U \otimes \Eins\otimes \cdots\otimes \Eins  \\
+ \dots + \ket{\mathbf{d}}\bra{\mathbf{d}}\otimes \Eins \otimes \cdots\otimes \Eins\otimes U,  
\end{multline}
where $\Eins$ is the identity applied to the respective register state. 
The gate $U_c$ is applied $M \in \ensuremath{\mathbbm N}$ times. 
If the ancilla is in the state $\ket{\mathbf{0}}$, the register states are acted upon by the identity. 
Otherwise, if the ancilla is in the state 
$\ket{\mathbf{j} \neq \mathbf{0}}$, the unitary $U$ acts on the $j$th register state $\ket{u_j}$, efficiently resulting in a phase $e^{i M \theta_j}$ imprinted on the ancilla state.
Therefore, after this step, the ancilla state is given by 
$\tfrac{1}{\sqrt{d+1}}\sum_{n=0}^d e^{i\left(M\theta_n+\phi_n\right)}\ket{\mathbf{n}}$,
where we defined $\theta_0=0$, while the state of the register qudits does not change since they are prepared eigenstates of $U$.
A second Hadamard gate finally maps the ancilla to the state
$\frac{1}{d+1}  \sum_{m,n=1}^d e^{i(M\theta_n + \phi_n+ \tfrac{2\pi mn}{d+1})}\ket{\mathbf{m}}$,
which is then measured in the computational basis. 
The probability of measuring the ancilla in one of the $d+1$ outcomes $o=0,\dots,d$ is
\begin{multline} \label{eq:main_probability}
P(o|\boldsymbol{\theta},\boldsymbol{\phi},M)=\frac{1}{(d+1)^2}\Big\{d+1 +2\sum_n\cos[M\theta_n+\beta_n]\\+2\sum_{n<m}\cos[M(\theta_n-\theta_m)+\gamma_{nm}]\Big\},
\end{multline}
where $\beta_n= \phi_n-\phi_0+\tfrac{2\pi n o}{(d+1)}$ and $\gamma_{nm}=\phi_n-\phi_m+\tfrac{2\pi (n-m) o}{(d+1)}$.

\subsection{Bayesian processing}

The algorithm consists of multiple rounds where, in the $k$th round, the basic quantum circuit of Fig.~\ref{fig1}(a) using $M=2^k$ ($k= 0, 1,\dots$) is repeatedly applied.
After setting the control phases $\boldsymbol{\phi}$, Bayes' theorem states that the posterior probability distribution corresponding to a measurement with result $o$ is 
\be \label{posterior}
P(\boldsymbol{\vartheta}|o,\boldsymbol{\phi},M) = \frac{P(o|\boldsymbol{\vartheta},\boldsymbol{\phi},M) P(\boldsymbol{\vartheta})}{\int \ud^d\boldsymbol{\vartheta} ~P(o|\boldsymbol{\vartheta},\boldsymbol{\phi},M) P(\boldsymbol{\vartheta})},
\ee
up to a normalization constant,
where $\boldsymbol{\vartheta}$ is a continuous $d$-dimensional variable, 
$P(o|\boldsymbol{\vartheta},\boldsymbol{\phi},M)$ is given in Eq.~(\ref{eq:main_probability}) and $P(\boldsymbol{\vartheta})$ is the Bayesian prior, 
which quantifies our knowledge about $\boldsymbol{\theta}$ before the measurement.
As initial condition (before the first measurement), we use a uniform prior $P(\boldsymbol{\vartheta}) = 1/(2\pi)^d$.
The posterior distribution $P(\boldsymbol{\vartheta}|o,\boldsymbol{\phi},M)$ expresses the degree of belief that 
$\boldsymbol{\theta} = \boldsymbol{\vartheta}$, 
where $\boldsymbol{\theta}$ is the true and unknown value of the vector parameter that we want to estimate, which is assumed constant while running the phase estimation algorithm.
Given the posterior distribution, we 
(i) update the Bayesian prior according to 
$P(\boldsymbol{\vartheta}|o,\boldsymbol{\phi},M)\mapsto P(\boldsymbol{\vartheta})$;
(ii) calculate $\bar{\theta}_j = \arg \int \ud^d \boldsymbol{\vartheta}~ e^{i\vartheta_j} P(\boldsymbol{\vartheta}|o,\boldsymbol{\phi},M)$
which provides an estimate of $\theta_j$; 
and (iii) calculate the Bayesian probability, $P_\mathrm{half}=\int_C \mathrm{d}^d P(\boldsymbol{\vartheta})$, of finding the true phase in the hypercube $C=\prod_j[\bar{\theta}_j-\pi/2^{k+1},\bar{\theta}_j+\pi/2^{k+1}]$ centered around $\bar{\boldsymbol{\theta}}$. 
If $P_\mathrm{half} \leq 1-\epsilon$, for some small $\epsilon$ that we indicate as the decision parameter, we set new control phases $\boldsymbol{\phi}$ and repeat the measurement, keeping the same $M=2^k$ as in the previous measurement.
If $P_\mathrm{half}>1-\epsilon$, we resrict the phase domain to the domain $C$, renormalize the Bayesian probability such that $\int_C \mathrm{d}^d P(\boldsymbol{\vartheta})=1$, and
proceed to the next $(k+1)$th round of the algorithm, namely, setting $M=2^{k+1}$. 
In Table~\ref{tab:algorithm1}, we show the pseudo code of the Bayesian algorithm. A detailed discussion of how many measurements are performed in each round is given in Appendix~\ref{ap:b}.

\begin{table}[b!]
\centering
\begin{tabular}{l}
\hline \hline
\noindent {\bf Input}: $\{\epsilon,k_\mathrm{max}\}$ \\
$P(\boldsymbol{\vartheta})=1/(2\pi)^d$\\
{\bf for } $k=0,\dots,k_\mathrm{max}-1$: \\
\hspace*{.4cm} $M=2^k$; $P_\mathrm{half}=0$ \\
\hspace*{.4cm} {\bf while } $P_\mathrm{half}<1-\epsilon$: \\
\hspace*{.8cm} $\boldsymbol{\phi} = \text{generate\textunderscore random}()$ \\
\hspace*{.8cm} $o = \mathrm{measurement}(M,\boldsymbol{\phi})$\\
\hspace*{.8cm} $P(\boldsymbol{\vartheta})= \text{Bay\textunderscore update} (P(\boldsymbol{\vartheta}),o,M,\boldsymbol{\phi})$\\
\hspace*{.8cm} $P_\mathrm{half}= \text{compute} (P(\boldsymbol{\vartheta}))$\\
\hspace*{.4cm} $P(\boldsymbol{\vartheta})=\text{cut \textunderscore grid}(P(\boldsymbol{\vartheta}))$\\
\hspace*{.4cm} $P(\boldsymbol{\vartheta})=\text{normalize}(P(\boldsymbol{\vartheta}))$\\
{\bf return}: $P(\boldsymbol{\vartheta})$\\
\hline \hline
\end{tabular}
\caption{The pseudocode of the Bayesian multiphase estimation algorithm. The initially flat phase distribution $P(\boldsymbol{\vartheta})$ is updated during $k_\mathrm{max}$ rounds of the algorithm. In the $k$th round, we set $M=2^k$ and perform measurements with random control phases $\boldsymbol{\phi}$ until we find $P_\mathrm{half}>1-\epsilon$, i.e., until the probability distribution is localized enough to cut the grid.}
\label{tab:algorithm1}
\end{table}

We observe that the use of random control phases $\boldsymbol{\phi} \in [0,2\pi]^d$, together with repeated measurements at fixed values of $M$, efficiently avoids the proliferation of tails in the Bayesian distribution, an effect that decreases the performance of the algorithm (see Appendix~\ref{ap:d}).
In the case of single-phase estimation, there are optimal adaptive methods to choose the control phase~\cite{Higgins2007b,Berry2009} that, however, cannot straightforwardly be generalized for the estimation of $d$ phases.
Furthermore, the use of random $\boldsymbol{\phi}$ avoids a (classical) computational slowdown associated with the calculation of optimal control phases. 
Alternatives to random control phases such as quasirandom sequences are discussed in Appendix~\ref{ap:d}.

A second important aspect is the cutting of the phase domain before increasing $M$. Cutting is necessary to guarantee a finite memory for storing the posterior distribution when running the algorithm with large $M$. Without cutting, there is an upper bound on $M$ imposed by the computer memory due to an increasing grid size.
Whenever $M$ is increased, our protocol instead allows to restrict the distribution to the hypercube $C$ of high probability, avoiding the need to increase the number of grid points.
Numerically, we find that a small finite grid size is advantageous because of small memory requirements and because it automatically performs a coarse graining (smoothening) of the Bayesian posterior.
The drawback is that each cutting of the phase domain produces an error probability $P_{\rm err}^{(d)}(\epsilon)$ of the true value $\boldsymbol{\theta}$ not being 
in the cut-out region $C$, thus resulting in a biased estimation.
Reduction of the error probability requires increasing the number of measurements in each round, and thus the total resources.  
In practice, from the numerical simulations reported in the following, we observe that
$P_{\rm err}^{(d)}(\epsilon) \sim \epsilon$, with a prefactor depending slightly on $d$.
The decreasing of the error probability $P_{\rm err}^{(d)}(\epsilon)$ requires increasing the number of measurements at each round as $O(\log(1/\epsilon))$ (see Appendix~\ref{ap:b}). 

\section{Results}\label{sec:simulation}
\subsection{Figure of merit and resource counting}

We quantify the performance of the algorithm by the $d \times d$ Bayesian covariance matrix 
\be \label{covariance}
\boldsymbol{V}_{ij} = 4 \int \ud^d \boldsymbol{\vartheta}~ 
\sin\frac{\vartheta_i - \bar{\theta}_i}{2}
\sin\frac{\vartheta_j - \bar{\theta}_j}{2} P(\boldsymbol{\vartheta}|o,\boldsymbol{\phi},M).
\ee
The diagonal entries of $\boldsymbol{V}$ correspond to the Holevo variance for small $V_{jj}$~\cite{Berry2000} while the off-diagonal elements
quantify correlations in the Bayesian distribution. Equation~(\ref{covariance}) gives the variance in 
the estimation of any linear combination of parameters $\boldsymbol{n} \cdot \boldsymbol{\theta}$ according to $V_{\boldsymbol{n} \cdot \boldsymbol{\theta}} =\sum_{i,j=1}^d n_i \boldsymbol{V}_{ij} n_j$,
where $\boldsymbol{n}$ is a $d$-dimensional real vector. 

As for resource counting, we follow the standard quantification of single parameter-estimation~\cite{Giovannetti2006,Higgins2007b,Berry2009}, namely, we count the number of applications of the controlled-$U$ gate, while assuming the existence of an oracle having access to $d$ eigenstates of $U$ and the possibility of generating controlled-$U^{M}$ gates. We want to emphasize why the generalized controlled gate $U_c$ corresponds a single application of the gate $U$: as one can see from Eq.~(\ref{CPG}), the application of the gate $U_c$ can only imprint the eigenphases $\theta_j$ (and not $d\times \theta_j$) on any initial state. This is in contrast to the application of $d$ unitaries $U$ (say, acting on the register states) that are able to imprint the phase $d\times \theta_j$ by properly choosing the initial state. Furthermore, considering the equivalence of the (multipass) Bayesian multiphase estimation algorithm to a protocol making use of N$00$N states [c.f. Sec.~\ref{sec:impl} and Eq.~(\ref{eq.GenNOON}], the $M$-fold application of $U_c$ corresponds to a single application of the multiparameter phase shift on a generalized N$00$N state with $M$ particles, counting as $M$ resources according to traditional resource counting in quantum metrology. 

Consequently, the total resources used up to the $K$th round of the algorithm are
$N_T = \sum_{k=0}^{K-1} m_k 2^k$, where $m_k$ is the number of measurements performed in the $k$th round, while the total number of measurements is $N_\mathrm{meas}=\sum_{k=0}^{K-1} m_k$ \cite{Wiebe2016}. 

\begin{figure*}[t!]
	\center
	\includegraphics[width = 1.7\columnwidth]{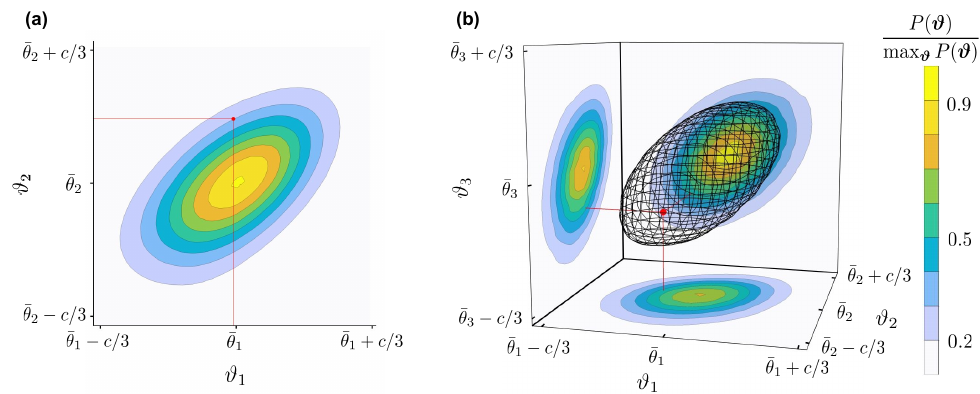}

	\caption{
	    Examples of the Bayesian probability distribution for (a) $d=2$ and (b) $d=3$, obtained after five rounds of the algorithm. 
	    The red point indicates $\boldsymbol{\theta}$. The phase domains are restricted to a fraction of the hypercube $C = \prod_{j=1}^d[\bar\theta_j - c, \bar\theta_j+c]$ with $c= \pi/2^{k}$ and $k=4$ in both cases. The color scale shows the Bayesian posterior $P(\boldsymbol{\vartheta})$ (a) and marginal Bayesian distributions (b) normalized to the maximum, ${\mathrm{max}}_{\boldsymbol{\vartheta}} P(\boldsymbol{\vartheta})$. 
	}
	\label{fig:Baydist}
\end{figure*}

\subsection{Sensitivity of the Bayesian algorithm in the noiseless case}

We perform extensive numerical simulations of the Bayesian multiphase estimation algorithm presented above.
In Fig.~\ref{fig:Baydist}, we show exemplary $d$-dimensional Bayesian posterior distributions for (a) $d=2$ and (b) $d=3$ parameters, after simulating the measurements up to the fifth round of the algorithm ($k=4$, $M=16$).
The Bayesian distributions are highly localized in a fraction of the hypercube $C$
and, as we discuss below, with a very high probability around the true $\boldsymbol{\theta}$. They are also clearly anisotropic
-- a direct consequence of the cross terms in Eq.~(\ref{eq:main_probability}) --
showing correlations between the different estimators $\bar\theta_j$. 

\begin{figure*}[t]
	\center
\includegraphics[width = 1.7\columnwidth]{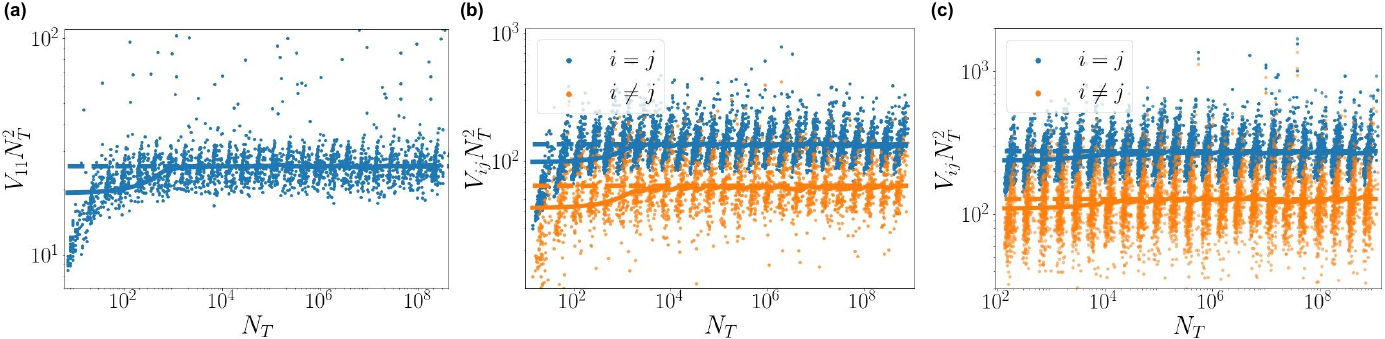}
	\caption{ Bayesian variances $V_{ij}$ ($i = j$) and covariances $V_{ij}$ ($i \neq j$) multiplied by $N_T^2$, for (a) $d=1$, (b) $d=2$ and (c) $d=3$. 
	Dots are results of numerical simulations and correspond to values calculated at the end of each round of the algorithm. Solid lines indicate averages converging to horizontal lines  is the average. It  converges to a horizontal lines $V_{ij} \sim N_T^{-2}$ for large $N_T$ (dashed, obtained from a fit) that provide the coefficients $C_H^{(d)}$ in Eqs.~(\ref{CHd1}-\ref{CHd3}). Here, we report results of $100$ simulations of the full algorithm up to $k=25$, estimating phases randomly chosen in $[0, 2\pi]^d$ with uniform distribution.
	The decision parameter is set to $\epsilon = 10^{-4}$.}
	\label{fig:Heisenberg}
\end{figure*}

\begin{figure*}[t]
	\center
	\includegraphics[width = 2\columnwidth]{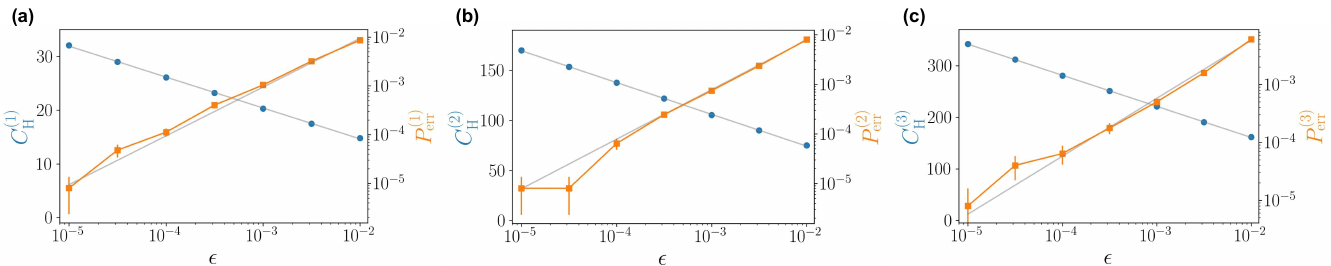}
	\caption{ 
	    The fitting parameter
	    $C^{(d)}_\mathrm{H}(\epsilon)$
	    of Eqs.~(\ref{CHd1})-(\ref{CHd3})
	    (blue dots) and the
	    error rate $P_\mathrm{err}^{(d)}(\epsilon)$ (orange squares) as a function of the decision parameter $\epsilon$, for (a) $d=1$, (b) $d=2$, and (c) $d=3$.
	    The solid gray lines are fits (see the text), error bars are estimated standard deviations of the error counts (see the text).
	    Averages are obtained by simulating the
	    algorithm up to $M= 2^{24}$ with $10^5$ repetitions for $d=1$ and $d=2$ and $5\times 10^4$ repetitions for $d=3$, respectively. 
	    }
	\label{fig:errorscaling}
\end{figure*}

In Fig.~\ref{fig:Heisenberg}, we show the results of the numerical simulation of the Bayesian covariance matrix, Eq.~\ref{covariance}.
Uniformly generated $\boldsymbol{\theta} \in [0,2\pi]^d$ are repeatedly estimated using $N_T \approx 10^9$, corresponding to 25 rounds, or about 500 measurements.
Fig.~\ref{fig:Heisenberg} shows the Bayesian variances $V_{jj}$ and covariances $V_{ij}$ for $i \neq j$ for (a) $d=1$, (b) $d=2$ and (c) $d=3$. 
The dots indicate numerical results obtained after each round of the algorithm (notice that there is no post-selection of optimal results). They spread around an average value (solid line) mainly due to fluctuations of the number $m_k$ of repeated measurements at each round. The horizontal dashed line is a numerical fit for $N_T\gg 1$, giving 
\be \label{CHd1}
\mathbf{V} = \frac{C^{(1)}_\mathrm{H}(\epsilon)}{N_T^2}, \qquad {\rm for}\,\, d=1,
\ee
\be \label{CHd2}
\mathbf{V} = \frac{C^{(2)}_\mathrm{H}(\epsilon)}{N_T^2}
\begin{pmatrix}
1 & 0.47 \\
0.47 & 1
\end{pmatrix}, \qquad {\rm for}\,\, d=2,
\ee
and 
\be \label{CHd3}
\mathbf = \frac{C^{(3)}_\mathrm{H}(\epsilon)}{N_T^2}
\begin{pmatrix}
1 & 0.45 & 0.45 \\
0.45 & 1 & 0.45 \\
0.45 & 0.45 & 1 
\end{pmatrix}, \qquad {\rm for}\,\, d=3,
\ee
where the fitting parameter $C^{(d)}_\mathrm{H}(\epsilon)$  depends on the decision parameter $\epsilon$. 
A fit of $C^{(d)}_\mathrm{H}$ as a function of $\epsilon$ (c.f. Fig.~\ref{fig:errorscaling}, blue dots), gives
$C^{(1)}_\mathrm{H} =  3.13 + 2.50 \ln{1/\epsilon}$ for $d=1$;
$C^{(2)}_\mathrm{H}= 10.8 + 13.8 \ln{1/\epsilon}$ for $d=2$; 
and  $C^{(3)}_\mathrm{H}= 40.1 + 26.2 \ln{1/\epsilon}$ for $d=3$.
In particular, 
the results of Fig.~\ref{fig:Heisenberg}, $C^{(2)}_\mathrm{H} \approx 138 $ and 
$C^{(3)}_\mathrm{H} \approx 281$, are obtained for $\epsilon=10^{-4}$. 
We thus have a scaling $O(N_T^{-2})$ of both diagonal and off-diagonal elements of the covariance matrix.
Also, the Bayesian covariance matrix scales exponentially with the number of measurements or, equivalently, with the number of ancilla qudits, as we show in Appendix~\ref{ap:a}.

Furthermore, we analyze the error rate $P_\mathrm{err}^{(d)}$ as a function of the decision parameter $\epsilon$.
More precisely, we calculate $P_\mathrm{err}^{(d)}$
as the probability 
-- per round --
that $\boldsymbol{\theta}\notin C$, namely, that the true value of the parameter lies outside the cut-out region $C$, resulting in a biased estimation of $\boldsymbol{\theta}$. $P_\mathrm{err}^{(d)}$ is estimated as follows. 
We observe numerically that, after a larger error probability in the first round, the error probability during the algorithm remains roughly constant (see Appendix~\ref{ap:c}). Therefore, by assuming a constant error probability per round, we slightly overestimate the asymptotic error probability per round. We simulate $N_\mathrm{sim}$ runs of the algorithm for different values of the decision parameter $\epsilon$, resulting in $N_\mathrm{err}$ errors (we count an error whenever the true value of $\boldsymbol{\theta}$ lies outside the hypercube $C$, $\boldsymbol{\theta}\notin C$). After running $k$ rounds, the probability of no error, $1-P_\mathrm{err,T}$, is related to the (constant) error probability per round, $P_\mathrm{err}^{(d)}$, by
\be 
1-P_\mathrm{err,T}^{(d)}=(1-P_\mathrm{err}^{(d)})^k. 
\ee
Using this relation and estimating $P_\mathrm{err,T}=N_\mathrm{err}/N_\mathrm{sim}$, we find an estimate for $P_\mathrm{err}^{(d)}$ (see Fig.~\ref{fig:errorscaling}, orange squares). The error bars are obtained by error propagation and the use of $\Delta P_\mathrm{err,T} = \sqrt{N_\mathrm{err}}/N_\mathrm{sim}$, assuming a Poisson distribution of error counts.

A fit of $P_\mathrm{err}^{(d)}(\epsilon)$ with the model $c \times \epsilon$ yields the error rate scalings $P_\mathrm{err}^{(1)} = 0.94 \epsilon$,
$P_\mathrm{err}^{(2)}= 0.78\epsilon$, and $P_\mathrm{err}^{(3)}= 0.58 \epsilon$, see Fig.~\ref{fig:errorscaling}. Note that a more accurate fit with the model $c_1\times \epsilon^{c_2}$ yields the parameters $c_1=0.57$ and $c_2=0.91$ for $d=1$, $c_1=0.94$ and $c_2=1.04$ for $d=2$,  and $c_1=0.64$ and $c_2=1.02$ for $d=3$. For a proper determination of the scaling of $P_\mathrm{err}^{(d)}(\epsilon)$, one should also perform more simulations. However, for our qualitative scaling comparison of parallel and sequential protocols, this rough estimation is sufficient.

In the case $d=1$, we can directly compare our results with existing single-phase estimation algorithms.
References~\cite{Higgins2007b,Berry2009} have studied ``backward'' algorithms that, in analogy to the semiclassical implementation of the inverse quantum Fourier transform~\cite{Griffiths1996}, are made of phase estimation rounds in a decreasing sequence 
$M=2^{K}, 2^{K-1}, ..., 1$.
Exploiting an optimized adaptive protocol, the  algorithm of Ref.~\cite{Higgins2007b}
reaches $V = C_s/N_T^2$ with $C_s = 23$.
The advantage of these approaches is that the error probability is negligible.
The drawback is that increasing the sensitivity by increasing $K$ requires to restart us the algorithm.
In contrast, our algorithm belongs to the class of ``forward'' algorithms~\cite{Higgins2009,Wiebe2016,Paesani2017,OBrien2019a,Pezze2020a}, where the phase sensitivity increases progressively as the protocol proceeds with, ideally, no upper bound in the number of rounds.
The Bayesian forward protocol of Ref.~\cite{Wiebe2016} approximates, at each round, the Bayesian distribution with a Gaussian function and reaches $C_s = 22$ (with a finite probability of biased estimation). 
As shown in Fig.~\ref{fig:Heisenberg}(a), we 
overcome the previous results, namely, we reach $C_H^{(1)} = 22$, at the price of an error probability of $P_\mathrm{err}^{(1)} = 5\times 10^{-4}$. 
Furthermore, our algorithm does not suffer from breakdown problems or an uncontrolled increase of error probability~\cite{Wiebe2016} due to the proliferation of tails in the Bayesian distribution.
The nonadaptive protocol of Ref.~\cite{Higgins2009} shows a negligible error probability at the price of a larger prefactor $C_s = 40.5$.

\subsection{Comparison between sequential and parallel multiphase estimation}

The estimation of $d$ eigenphases can be performed either following a parallel protocol using an $(d+1)$-dimensional ancilla qudit, as illustrated above,
or following a sequential protocol where each phase is estimated separately using ancilla qubits [see Fig.~\ref{fig1}(b)].
In the latter case, any linear combination of $d$ eigenphases is estimated with sensitivity 
\be
V _{\boldsymbol{n} \cdot \boldsymbol{\theta}_{\rm seq}} = \sum_{j=1}^d n_j^2 V_j = \frac{C_H^{(1)} d^2 \sum_{j=1}^d n_j^2}{N_T^2},
\ee
where $V_j$ is the sensitivity of estimating $\theta_j$ in a single-parameter estimation algorithm and $N_T$ is the total number of resources used in the multiphase estimation. We assume here that all phases are estimated using the same resources (note that an optimization of resources, depending on the values of $n_j$, is also possible).

Compared to the sequential protocol, the parallel protocol offers two key advantages.
(1) The estimation of multiple phases simultaneously
might be necessary, or might prove particularly efficient, in systems and applications where spatial or temporal constraints (e.g., for time-varying signals) prevent the estimations of single phases independently. 
(2) By exploiting correlations in the Bayesian multiphase distribution, certain combinations of the eigenphases can be estimated 
with higher sensitivity in the parallel case compared to the sequential case, for a fixed error rate $P_{\rm err}$ in both cases.
Explicitly, in the case $d=2$, the sequential strategy estimates $\theta_1$ and $\theta_2$ in a single-phase estimation and then infers $\theta_1-\theta_2$ with sensitivity
$V_{\theta_1 - \theta_2} = 8C^{(1)}_\mathrm{H}(P_{\rm err})/N_T^2$ with $C^{(1)}_\mathrm{H}(P_{\rm err})=3.13 + 2.50 \ln{0.94/P_{\rm err}}$. On the other hand, the parallel estimation uses $V_{\theta_1 - \theta_2} = 2(1-0.47)C^{(2)}_\mathrm{H}(P_{\rm err})/N_T^2$ with $C^{(2)}_\mathrm{H}(P_{\rm err})=10.8 + 13.8 \ln{0.78/P_{\rm err}}$.
Fixing, for instance, $P_{\rm err} = 10^{-3}$, the sequential protocol yields $V_{\theta_1 - \theta_2} = 162/N_T^2$, while the parallel protocol scales as  
$V_{\theta_1 - \theta_2} =109/N_T^2$. 
Analogously, for $d=3$ and $P_{\rm err} = 10^{-3}$, the sequential scheme measures all phase differences with $V_{\theta_j - \theta_k} = 364/N_T^2$, compared to $V_{\theta_j - \theta_k} = 227/N_T^2$ for the parallel protocol. 

\subsection{Sensitivity of the Bayesian algorithm in presence of noise}

In the following, we extend the Bayesian multiphase estimation algorithm in the presence of noise. In particular, we assume
that each phase imprinting is accompanied by noise characterized by a decoherence rate $\Gamma_j$ for the $j$th phase.
In particular, we generalize the qubit phase damping channel~\cite{Nielsen2010} to the qudit case. Formally, if the ancilla state is given by the density matrix $\rho$, the phase damping is described by the channel $E(\rho)=\sum_{j=0}^d K_j \rho K_j$, where the Kraus operators are defined as  $(K_0)_{mn}=e^{-\Gamma_m M}\delta_{mn}$ ($\Gamma_0=0$) 
and $(K_j)_{mn}=\sqrt{1-e^{-2\Gamma_j M}}\delta_{mj}\delta_{nj}$ for $j=1,\dots, d$.
In the case $d=1$, this model results in a decay of the measurement probabilities to white noise, a model commonly considered in single-phase estimation~\cite{Wiebe2016}. 
In the optical implementation (cf. Sec.~\ref{sec:impl}), the noise model can be understood as a mode-dependent dephasing during each phase imprinting.

The algorithm in the presence of noise follows the noiseless one with setting $M=\min_j[2^k,1/\Gamma_j]$ in the $k$th round. In other words, we stop increasing $M$ as soon as the noise becomes dominant. After this point, the covariance matrix $\boldsymbol{V}$ follows a shot-noise scaling ($V_{ij}\sim N_T^{-1}$). The pseudocode of the Bayesian algorithm in the presence of noise is shown in Table~\ref{tab:algorithm2}. 

\begin{table}[h!]
\centering
\begin{tabular}{l}
\hline \hline
\noindent {\bf Input}: $\{\epsilon,k_\mathrm{max},\Gamma_j\}$ \\
$P(\boldsymbol{\vartheta})=1/(2\pi)^d$\\
{\bf for } $k=0,\dots,k_\mathrm{max}-1$: \\
\hspace*{.4cm} $M=\min_j[2^k,1/\Gamma_j]$; $P_\mathrm{half}=0$ \\
\hspace*{.4cm} {\bf while } $P_\mathrm{half}<1-\epsilon$: \\
\hspace*{.8cm} $\boldsymbol{\phi} = \text{generate\textunderscore random}()$ \\
\hspace*{.8cm} $o = \mathrm{measurement}(M,\boldsymbol{\phi})$\\
\hspace*{.8cm} $P(\boldsymbol{\vartheta})= \text{Bay\textunderscore update} (P(\boldsymbol{\vartheta}),o,M,\boldsymbol{\phi})$\\
\hspace*{.8cm} $P_\mathrm{half}= \text{compute} (P(\boldsymbol{\vartheta}))$\\
\hspace*{.4cm} $P(\boldsymbol{\vartheta})=\text{cut \textunderscore grid}(P(\boldsymbol{\vartheta}))$\\
\hspace*{.4cm} $P(\boldsymbol{\vartheta})=\text{normalize}(P(\boldsymbol{\vartheta}))$\\
{\bf return}: $P(\boldsymbol{\vartheta})$\\
\hline \hline
\end{tabular}
\caption{The pseudocode of the Bayesian multiphase estimation algorithm in the presence of noise. The measurement strategy is similar to the one in the noiseless case (cf. Table~\ref{tab:algorithm1}). Here, in the $k$th round, we choose $M=\min_j[2^k,1/\Gamma_j]$ dependent on the different dephasing rates $\Gamma_j$.}
\label{tab:algorithm2}
\end{table}

\begin{figure}[t!]
	\center
\includegraphics[width = 0.9\columnwidth]{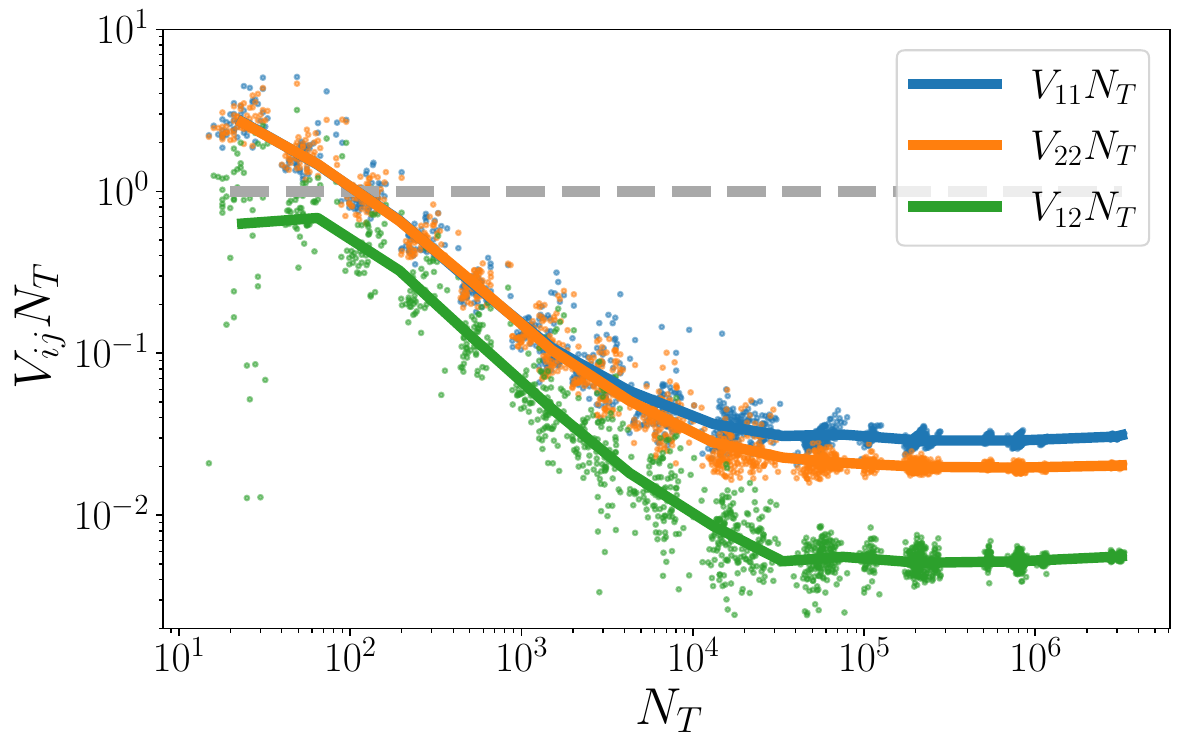}
	\caption{The scaling of the Bayesian 
	variances $V_{ij}$ ($i = j$) and covariances $V_{ij}$ ($i \neq j$)
	with the total number of resources $N_T$ in the presence of dephasing. The dots indicate results of numerical simulations after each round and the lines correspond to average values. The dashed line is the shot-noise limit  $N_T^{-1}$. Here,
	$d=2$, $\Gamma_1=0.02$, $\Gamma_2=0.01$ and $\epsilon=10^{-4}$.}
	\label{fig:noise}
\end{figure}

In Fig.~\ref{fig:noise}, we show results of numerical simulations, plotting $N_T V_{ij}$ as a function of $N_T$, for $d=2$ phases with $\Gamma_1=0.02$ and $\Gamma_2=0.01$. 
The change of scaling of $\boldsymbol{V}$ around $N_T = 10^4$ is evident, approaching a constant value for sufficiently large $N_T$. 
In this regime, the estimation is still overcoming the shot-noise limit $V_{\theta_j}=N_T^{-1}$ (dashed line in Fig.~\ref{fig:noise}).
Note that since the strong correlation in the noiseless case stems from the cross terms of Eq.~\ref{eq:main_probability}, the correlations now decay with a larger rate $\Gamma_1+\Gamma_2$. 
We also mention that an asymmetric noise can be balanced in
the estimation algorithm by performing each measurement with an $M$-dependent unbalanced Hadamard gate $\tilde H_{d+1}$ instead of the first Hadamard gate.

\section{Optical implementation}\label{sec:impl}

Above, we focus on the parallel estimation of $d$ eigenphases of a unitary matrix: a task that can find possible applications in a wide variety of quantum computing tasks based on QPE.
In addition, here we show that our methods can be straightforwardly applied to multiparameter estimation problems in quantum optical interferometry.
In particular, we consider the parallel estimation of $d$ optical phase shifts using a generalized Mach-Zehnder interferometer setup.

\begin{figure}[t!]
	\center
	\includegraphics[width = 0.9\columnwidth]{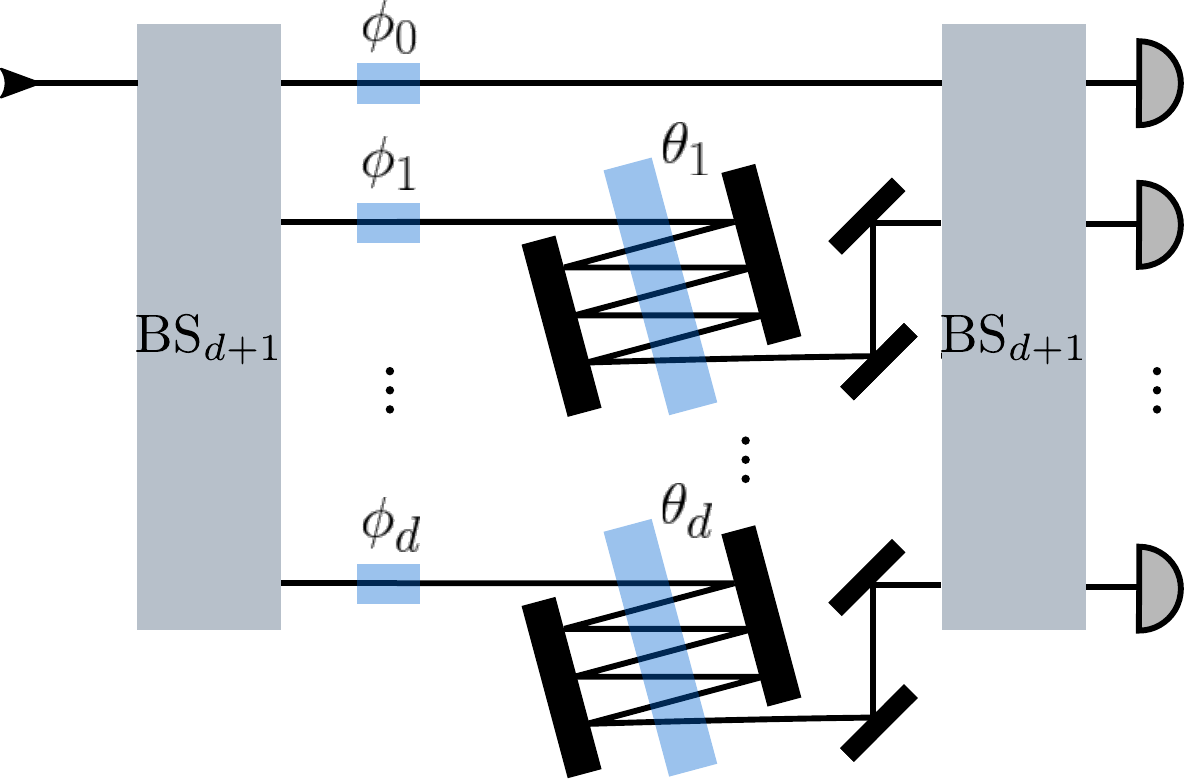}
	\caption{
	    The optical implementation of the parallel multiphase estimation algorithm. A single photon enters a multiarm interferometer composed of two $(d+1)$-port beam-splitters (${\rm BS}_{d+1}$) that enclose a control ($\phi_0, ..., \phi_d$) and an unknown $(\theta_1, ..., \theta_d)$ optical phase shift in each mode. The phase shifts $\theta_j$ are defined with respect to a reference phase $\theta_0$. The photon is finally measured in the outcoming modes.}
	\label{fig:implementation}
\end{figure}

The protocol can be realized with single photons and standard optical elements and is thus accessible by current state-of-the-art technology~\cite{Spagnolo2012,Polino2019,Polino2020} (see Fig.~\ref{fig:implementation}). 
The extended Hadamard gate is realized by a multiport beam splitter $\mathrm{BS}_{d+1}$ ~\cite{Z-dotukowski1997,Spagnolo2012},
such as a tritter for $d=2$ or a quarter for $d=3$, recently implemented with integrated optical circuits~\cite{Spagnolo2012, Polino2019, Polino2020, Valeri2020}, that can be
also realized with a cascade of 50:50 beam 
splitters~\cite{Guo2020}.
For instance, in the case $d=2$, the unitary transformation of a tritter is given by~\cite{Spagnolo2012, Spagnolo2013} 
\be
U_\mathrm{BS,3} = \frac{1}{\sqrt{3}}\begin{pmatrix}
1 & 1 & 1 \\
1 & e^{i2\pi/3} & e^{i4\pi/3} \\
1 & e^{i4\pi/3} & e^{i2\pi/3}
\end{pmatrix}
\ee 
which corresponds precisely to $H_{3}$ in Sec.~\ref{sec:alogorithm}.
In the case $d=3$, the corresponding optical element is called a quarter and is parametrized by a implementation-dependent parameter $\gamma$ \cite{Z-dotukowski1997},
\be
U_\mathrm{BS,4}(\gamma) = \frac{1}{2}\begin{pmatrix}
1 & 1 & 1 & 1\\
1 & e^{i\gamma} & -1 & -e^{i\gamma}\\
1 & -1 & 1 & -1\\
1 & -e^{i\gamma} & -1 & e^{i\gamma}
\end{pmatrix}.
\ee
For $\gamma=\pi/2$, this corresponds exactly to $H_4$ in Sec.~\ref{sec:alogorithm}, while for other $\gamma$, the final probability has to be marginally adjusted.

The different phases $\theta_1, ..., \theta_d$ and the (random) control phases $\phi_0, ..., \phi_d$ are imprinted on the different optical modes. The phase shift on mode $j$ is implemented by the transformation $\exp[i( M \theta_j+\phi_j)n_j]$, where $n_j$ is the number operator for the $j$th mode.
The phase ``amplification'' $M \times \theta_j$ is obtained by $M$ passes of the phase shift $\theta_j$, as demonstrated experimentally for single-phase estimation in Ref.~\cite{Higgins2007b}.
Note that all $d$ phases are defined with respect to an unknown reference phase $\theta_0$ in the zeroth arm.
Finally, after a second multiport beam splitter, the photon is
 measured in one of the $d+1$ output modes, resulting in statistics described by Eq.~\eqref{eq:main_probability}. 
As discussed above, the optical scheme of Fig.~\ref{fig:implementation} would reach a multiphase estimation variance with optimal scaling $O(N_T^{-2})$ with respect to the total resources $N_T$ (quantified here as the total number of applications of the phase shift $\boldsymbol{\theta}$), and exponentially with the number $n_{\rm phot}$ of single photons used (cf. Appendix~\ref{ap:a}).

Finally, we mention that, in principle, the protocol can be equivalently implemented with multimode N$00$N-like states~\cite{Humphreys2013,Proctor2018,Ge2018,Gessner2018b,Zhang2018}
\be \label{eq.GenNOON}
\ket{\psi} = \frac{\ket{M, 0, ..., 0}+ \ket{0, M, 0 ..., 0} + \ket{0, ...,0, M}}{\sqrt{d+1}}.
\ee
The different phases in each mode are imprinted by a single application of the phase shifts, namely, $\exp[i(\theta_j+\phi_j)n_j]$. Projecting the final state on the set of states 
\begin{multline}
\ket{\psi_o} = \frac{1}{\sqrt{d+1}} \bigg( \ket{M, 0, ..., 0} + e^{i\tfrac{2\pi o}{d+1}}\ket{0, M, ..., 0}\\ + \dots  +  e^{i\tfrac{2\pi d o}{d+1}}\ket{0, 0, ..., M} \bigg),
\end{multline}
where $o=0,\dots,d$ labels the measurement outcome,
results in the probabilities of Eq.~\eqref{eq:main_probability} (redefining the random control phases $\phi_j \rightarrow \phi_j/M$).
In this scheme, the phases $M\times \theta_j$ are imprinted by a single application of the phase shift $\theta_j$ but using $M$ photons.
Regarding resources, the two optical schemes are equivalent~\cite{Higgins2007b,Giovannetti2006,Dowling2008}.
Although proposed in Ref.~\cite{Humphreys2013} and largely studied as a benchmark in the multiphase estimation literature~\cite{Proctor2018,Ge2018,Gessner2018b}, the generalized NOON state of Eq.~(\ref{eq.GenNOON}) has not been realized experimentally so far.

\section{Conclusions}\label{sec:conclusion}
To summarize, we have presented a Bayesian quantum algorithm for the simultaneous estimation of $d$ arbitrary phases $\bs{\theta} \in [0,2\pi]^d$ with covariance matrix elements scaling as $O(N_T^{-2})$ with respect to the total number of resources $N_T$.
The running of the algorithm requires the support of a classical memory to store the Bayesian distribution. This is necessary to calculate variances, confidence intervals, and estimators, and to monitor the process of the algorithm.
Our algorithm is well suited for the estimation of a few parameters, realizable with an optical device \cite{Spagnolo2012,Polino2019,Polino2020}.
For larger values of $d$, other cutting strategies that do not scale with $d$ can be considered, such as Gaussian approximations of the Bayesian distribution~\cite{Wiebe2016} or adaptive updates of the grid points using Monte Carlo methods~\cite{Granade2012,Valeri2020}. However, the impact of these methods on the probability of obtaining biased estimations has to be carefully analyzed.

In the context of multiparameter quantum metrology, our protocol 
provides a viable strategy to estimate arbitrary phases $\theta_j$ and their linear combinations $\bs{\theta}\cdot \bs{n}$ with $\Delta^{2} (\bs{\theta}\cdot \bs{n}) \sim 1/N_T^{2}$, for fixed resources $N_T \gg 1$. Furthermore, 
our analysis highlights that correlations in the Bayesian distribution (only accessible by the parallel scheme) can be exploited for the estimation of certain combinations of phases with a sensitivity overcoming that achievable by sequential (single-parameter) estimation strategies. Finally, an optical version of the algorithm using single photons and linear optical elements can be implemented in state-of-the-art experiments~\cite{Spagnolo2012,Polino2019,Polino2020}. Equivalently, the algorithm can be implemented using generalized N00N-states~\cite{Humphreys2013,Proctor2018,Ge2018,Gessner2018b}.
Recently, a multiround protocol using entangled qubits to measure a specific linear combination of multiple phases (at an optimal point) has been realized \cite{Liu2021,Zhao2021}. In contrast, our protocol is able to measure all linear combinations of phases simultaneously.

\section*{Acknowledgments}
We acknowledge financial support from the European Union's Horizon 2020 research and innovation programme - Qombs Project, FET Flagship on Quantum Technologies Grant No. 820419. 

\appendix
\section*{Appendix}

\section{Number of measurements per round}\label{ap:b}

\begin{figure}[b]
	\center
	\includegraphics[width = .99\columnwidth]{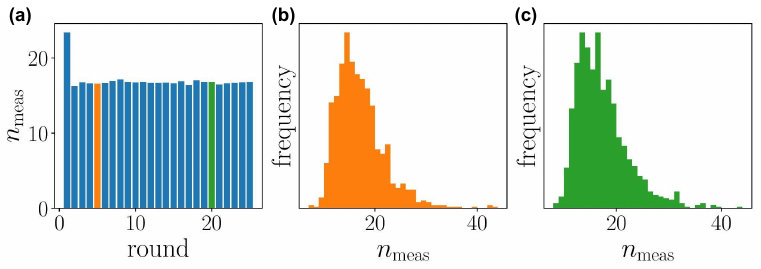}
	\caption{
	(a) The average number of measurements per round, $n_\mathrm{meas}$, used in the different rounds of the algorithm simulated for $1000$ runs. (b-c) The distribtuion of the number of measurements $n_\mathrm{meas}$ used in (b) the fifth round and in (c) the $20$th round of the algorithm. $d=2$, $\epsilon=10^{-4}$.
	}
	\label{fig:meas_dist}
\end{figure}
We now discuss the distribution of the number of measurements in the different rounds of the phase estimation algorithm. 
Here, $n_\mathrm{meas}(k)$ is the number of measurements used in the $k$th round when averaged over many runs of the algorithm. 
It is related to the average total number of measurements after $K$ rounds, $\langle N_\mathrm{meas}(K)\rangle$, as $\langle N_\mathrm{meas}(K)\rangle=\sum_{k=0}^K n_\mathrm{meas}(k)$.
In Fig.~\ref{fig:meas_dist}(a), we show $n_\mathrm{meas}$ depending on the round for $1000$ simulations of the algorithm until $M=2^{24}$ for $d=2$ and $\epsilon=10^{-4}$. 
We see that after the first round, a constant number of measurements is performed on average per round. In the first round, more measurements are performed because the initial flat Bayesian distribution has to become sufficiently localized before the algorithm switches to the second round. 

In Figs.~\ref{fig:meas_dist} (b) and ~\ref{fig:meas_dist} (c), we show the distribution of the number of measurements $n_\mathrm{meas}$ in (b) the fifth round and in (c) the $20$th round for the simulation of $1000$ runs used for Fig.~\ref{fig:meas_dist} (a) (indicated in various colors). We observe that the distribution of $n_\mathrm{meas}$ is similar for the different rounds and reason that, due to the tail of the distribution, the use of a constant number of measurements per round might lead to an increased error if chosen too small or to a large constant $C_\mathrm{H}^{(d)}$ of the Heisenberg scaling if chosen too large. 

\begin{figure}[t]
	\center
	\includegraphics[width = .8\columnwidth]{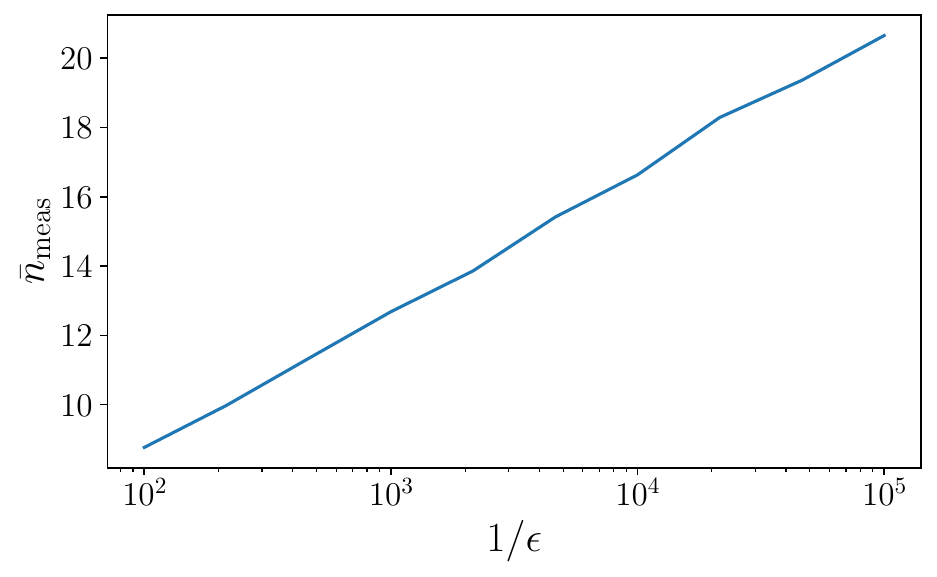}
	\caption{
	The scaling of the average number of measurements per round, $\bar n _\mathrm{meas}$, (excluding the first round) used in the algorithm depending on the decision parameter $\epsilon$. For ten different values of $\epsilon$, we simulate $100$ runs of the algorithm and average the results. $d=2$.
	}
	\label{fig:meas_over_epsilon}
\end{figure}

Having checked that after the first measurement, the number of measurements per round is constant on average, we now check the scaling of this average number $\bar{n}_\mathrm{meas}$ (excluding the first round) depending on the decision parameter $\epsilon$. In Fig.~\ref{fig:meas_over_epsilon}, we show $\bar{n}_\mathrm{meas}$ for different values of $\epsilon$ averaged over the different rounds of $100$ runs of the algorithm for each $\epsilon$ and $d=2$.
We see that $\bar{n}_\mathrm{meas}$ scales as $O(\ln 1/\epsilon)$ (or, equivalently, $O(\ln 1/P_\mathrm{err}^{(d)})$), a scaling that is also observed in the Kitaev phase estimation algorithm \cite{Kitaev1996}. In Ref. \cite{Kitaev1996}, the unknown phase is estimated (with high probability) up to an additive error $\varepsilon$  using $O(\ln 1/\varepsilon)$ ancilla qubits. 

\section{Choice of the control phases during the algorithm}\label{ap:d}

\begin{figure*}[t]
	\center
	\includegraphics[width = 1.9\columnwidth]{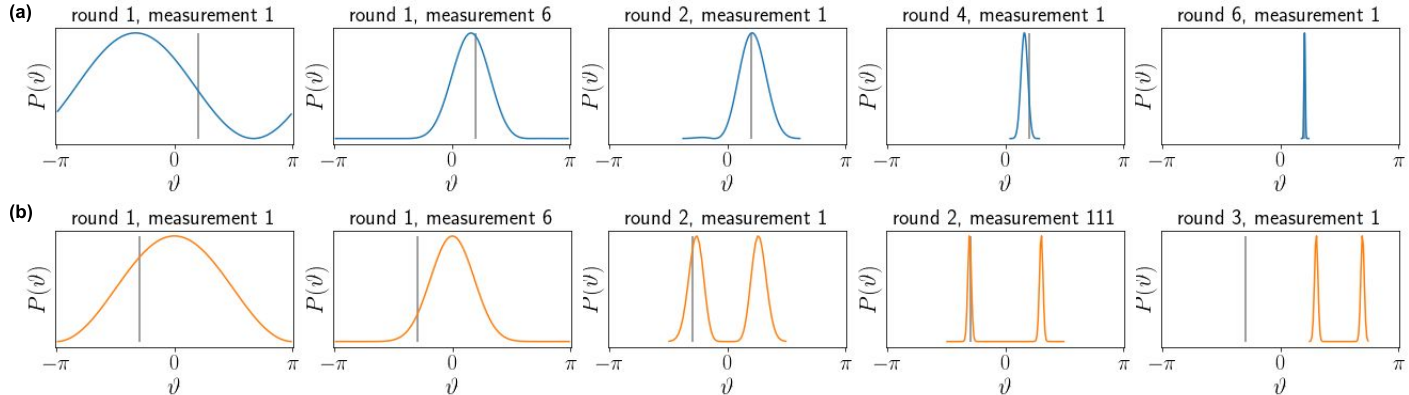}
	\caption{
	The Bayesian distributions during the algorithm for $d=1$ and $\epsilon=10^{-4}$ for (a) random control phases and (b) a constant (zero) control phase (b).
	}
	\label{fig:choices}
\end{figure*}

Here, we will discuss the choice of the random control phases during the algorithm. First, we discuss the case $d=1$ for simplicity. If we do not vary the control phase (without loss of generality, consider $\phi=0$), the Bayesian update results in products of $\cos^2 \theta/2$ and $\sin^2 \theta/2$ with multiplicities depending on the different measurement outcomes. This will eventually lead to a double-peak structure in the Bayesian distribution. Measurements with a larger $M$ in later rounds are not sufficient to resolve the two peaks, leading finally to an erroneous cutting and to a biased estimation. This behaviour is shown in Fig.~\ref{fig:choices}(b). It is known that in the case of $d=1$, this obstacle can be solved by alternating the control phase between $\phi_{2n}=0$ and $\phi_{2n+1}=\pi/2$ for the $n$th measurement in each round (see, e.g., Ref. \cite{Higgins2009}). The same effect is reached by using random (but known) control phases in each measurement. 

In the case $d=2$, an alternative to using random control phases is to set the control phases $\boldsymbol{\phi}$ in the $n$th measurement to $(0,0)$ if $n \mod 4 = 1$, $(\pi/2,0)$ if $n \mod 4 = 2$, $(0,\pi/2)$ if $n \mod 4 = 3$ and $(\pi/2,\pi/2)$ if $n \mod 4 = 0$. For $\epsilon=10^{-3}$, this leads to $C_\mathrm{H}^{(2)} = 113 $ and $P_\mathrm{err}^{(2)} = 7.1\times 10^{-4}$, compared to $C_\mathrm{H}^{(2)} = 105$ and $P_\mathrm{err}^{(2)} = 7.7\times 10^{-4}$ for random phases. 

A different approach is to use so-called quasi-random sequences \cite{Niederreiter1978}. Quasi-random sequences are deterministic sequences of tuples $\boldsymbol{x}_n$ that densely cover $[0,2\pi]^2$ for $n\to\infty$ and appear random but, at the same time, cover $[0,2\pi]^2$ more evenly than random numbers during the convergence. Using, for example, the Halton sequence in two dimensions as control phases, we observe $C_\mathrm{H}^{(2)} = 101 $ and $P_\mathrm{err}^{(2)} = 8.5\times 10^{-4}$ for $\epsilon=10^{-3}$. 
We see that the alternatives of random control phases discussed here yield very similar results to random ones. A minor advantage could be achieved by using quasi-random control phases.  
For $d>2$, similar choices of control phases can be made. 

\section{Scaling of the precision with the number of measurements}\label{ap:a}

\begin{figure}[t]
	\center
	\includegraphics[width = .8\columnwidth]{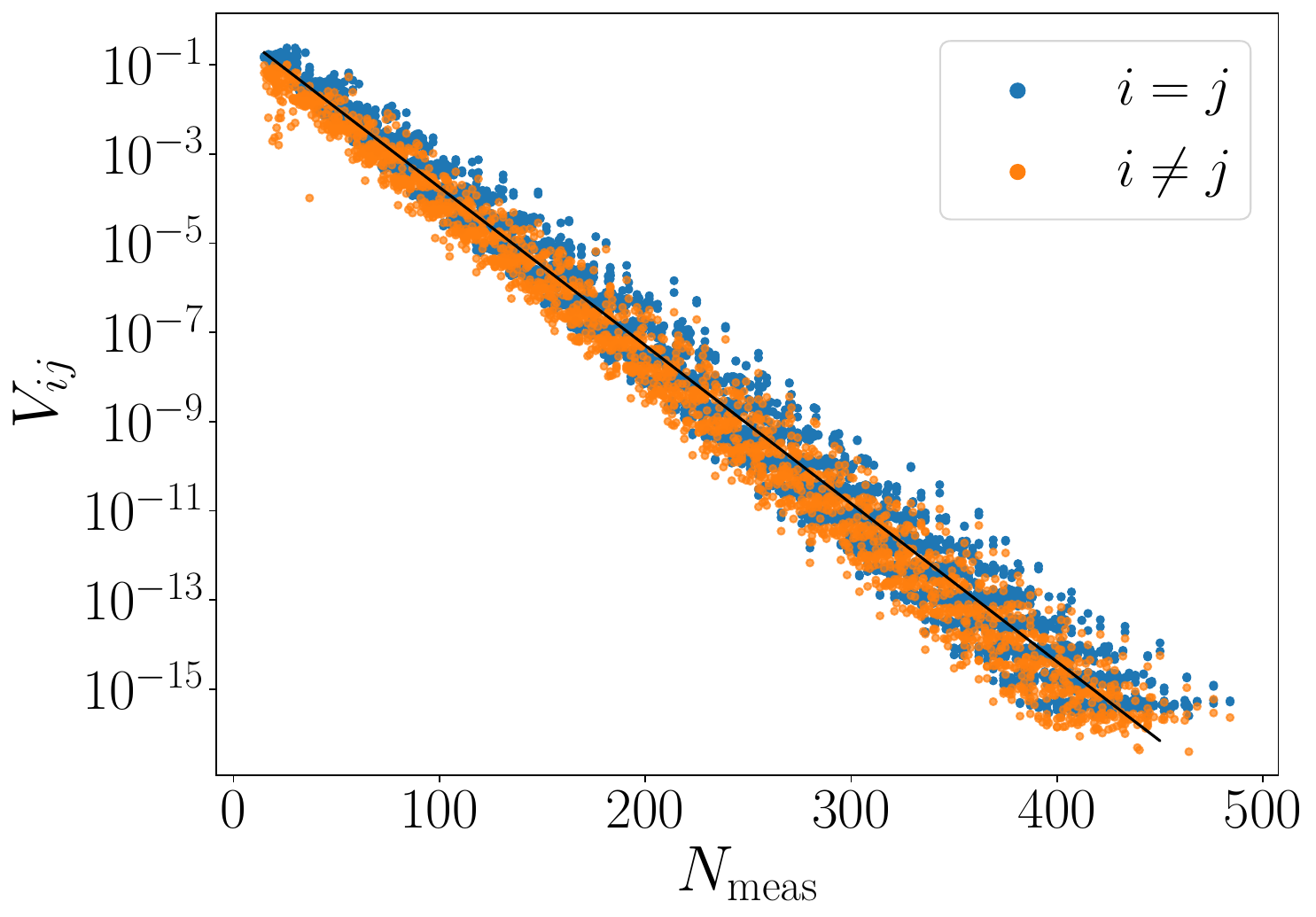}
	\caption{
	The scaling of the components $V_{ij}$ of the covariance matrix $\mathbf{V}$ as a function of the total number of measurements $N_\mathrm{meas}$. The dots are results of $100$ runs of the multiphase estimation algorithm for $d=2$ for $M\leq 2^{24}$. The solid line is a fit of the variances $V_{jj}$, see text. $\epsilon=10^{-4}$.
	}
	\label{fig:expscaling}
\end{figure}

We study the exponential scaling of the Bayesian covariance matrix $\boldsymbol{V}$ with the number of measurements, $N_\mathrm{meas}$. In Fig.~\ref{fig:expscaling}, we plot the scaling of the different components $V_{ij}$ as a function of $N_\mathrm{meas}$ for the multiphase estimation algorithm estimating $d=2$ phases and a decision parameter $\epsilon=10^{-4}$. Shown are results from $200$ realizations of the algorithm up to $M=2^{24}$ (dots) and an exponential fit $V_{jj}(N_\mathrm{meas})=0.61\exp[- 0.082 N_\mathrm{meas}]$ (solid line). An analogous scaling can be observed for $d=1$ and $d=3$. In the implementation of the algorithm using controlled-$U$ operations, $N_\mathrm{meas}$ corresponds to the total number of ancilla qubits that have to be prepared and measured. For the optical implementation using single photons, $N_\mathrm{meas}$ is the total number of photons used in the algorithm and for the optical implementation using generalized N00N states, $N_\mathrm{meas}$ is the total amount of applications of the unknown phase shifts. 

\section{Analysis of error probability per round}\label{ap:c}

\begin{figure}[t]
	\center
	\includegraphics[width = .99\columnwidth]{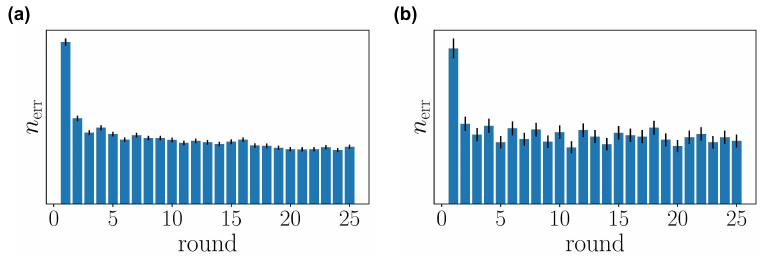}
	\caption{
	The distribution of errors per round for $d=1$ and a decision parameter (a) $\epsilon=10^{-2}$ and (b) $\epsilon=10^{-3}$. Each plot shows a histogram of errors after the simulations of $10^6$ runs. The error bars ($\sqrt{n_\mathrm{err}}$) are the estimated standard deviation assuming a Poisson distribution of error counts.  
	}
	\label{fig:zero_dists}
\end{figure}

In Fig.~\ref{fig:zero_dists}, we show simulations of the error probability for $10^6$ runs of the algorithm with $d=1$, using decision parameters (a) $\epsilon=10^{-2}$ and (b) $\epsilon=10^{-3}$. We observe that after an increased error probability in the first round, the error probability per round remains roughly constant and even decreases slightly for later rounds. Therefore, the error analysis in Sec.~\ref{sec:simulation}, which counts all errors that occur after $25$ rounds of the algorithm and assumes a constant error probability in each round, represents an overestimation of the real asymptotic error rate of the algorithm.

\end{document}